\documentclass[12pt,preprint,referee]{elsarticle}
\usepackage{graphicx}  
\usepackage{dcolumn}   
\usepackage{bm}
\newcommand{\mnras}{MNRAS}             
\newcommand{\aj}{AJ}                   
\newcommand{\dd}{{\rm d}}

\begin{document}

\begin{frontmatter}
\title{Different thermodynamics of self-gravitating systems and discussions for some observations and simulations}
 \author{Dong-Biao Kang$^{1}$\thanks{E-mail:
billkang@itp.ac.cn}\\
$^{1}$ Key Laboratory of Frontiers in Theoretical Physics, Institute of Theoretical Physics, Chinese Academy of Science, Beijing100190, China}
\date{Received  / Accepted}
\maketitle
\begin{abstract}
Our previous works have shown the statistical mechanics of self-gravitating system. In this paper, we will show its thermodynamics and compare our results with observations and simulations. We propose that our statistical mechanics can be based on ergodicity breaking and Boltzmann entropy, and its assumptions do not contradict with the reality. With the principle of statistical mechanics, we will show our definition of temperature and then the capacity of self-gravitating systems. We find that the gravothermal catastrophe may be a special case of our theory. Our results also provide new explanations for the density profiles of observations and numerical simulations, especially we think that the non-universal density distribution in the simulations of dissipationless collapse is not caused by the different initial collapse factor. We will also discuss about the core-cusp problem.
\end{abstract}
\end{frontmatter}

\section{Introduction}

\label{1}
Statistical mechanics of systems with long-range interactions presently has not been generally considered to be well established, because thermodynamics of long-range system is very different: the energy is non-additive; the entropy needs not to be a concave function of energy, so the specific heat may be negative in the microcanonical ensembles \citep{ant62,lyn68,thir06}; then the microcanonical ensembles and canonical ensembles may be not equivalent; the ergodicity which is the fundamental assumption of statistical mechanics may be broken \citep{mukamel08}; and many others.

However, our latest studies provide different understandings of the long-range statistical mechanics. \cite{hk10} preliminarily studies the fluid entropy and proposes a self-gravitating system's entropy principle which is different from the well known principle of maximum entropy. \cite{kh11} completes the variation process of entropy and confirms that this entropy solves the problem that people always obtain the infinite mass and energy when using the Boltzmann-Gibbs entropy. Now it is necessary to show the thermodynamics corresponding to our statistical mechanics.

Besides, presently there have been many observations and increasing numerical simulations of some astronomical systems which can be approximately considered to be self-gravitating. While there are many observations of globular clusters and elliptical galaxies, there are few simulations of dissipationless collapse of galaxies which will interest us here. More current simulations are cosmological simulations which have been successful to be consistent with the universe's large scale structure predicted by the $\Lambda$CDM model, but there seems to be some contradictions with observations at the galaxy scale, such as that current cosmological simulations provide us a NFW density profile \citep{nfw}
\begin{equation}
\label{nfw}
\rho(r)=\frac{4\rho_s}{r/r_s(1+r/r_s)^2},
\end{equation}
which has a central cusp and is not consistent with the measurements of Low Surface Brightness galaxies \citep{deblok09} that show us a central core. Other two inconsistencies between simulations and observations include the missing satellite and the problem about the distribution of angular momentum \citep{moore99}. In this letter we will also discuss some of these results.
\section{Equation of state}
\label{2}
\begin{figure}
{\includegraphics[width=\columnwidth]{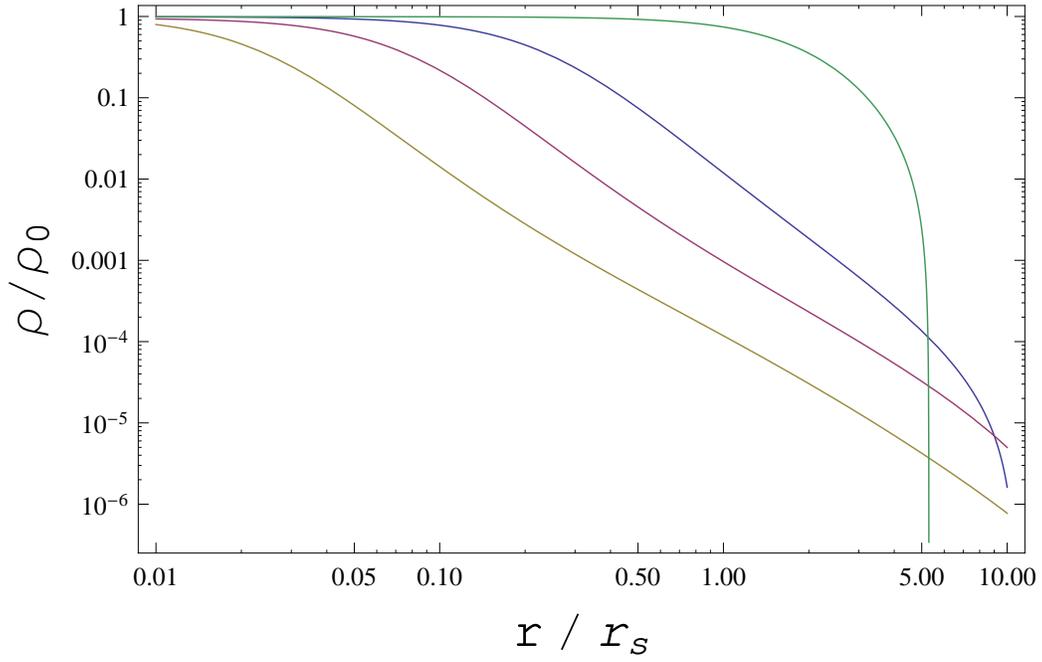}}
\label{density}
\caption{The density profiles with different values of $\beta$ and fixed $\alpha$. $\rho_0$ is the central density, $r_s$ is a characteristic scale, and we assume that these two quantities have been given. With the increase of $\beta$ from $0$ to $\infty$, the region of the central core becomes small and the density profile tends to the singular isothermal sphere.}
\end{figure}

We first summarize some of our previous works with more observational evidences. Our attentions are restricted to the spherical self-gravitating system. It is known that if we insist on the ergodicity and the principle of maximum principle, we will get the isothermal solution which gives infinite mass and energy \citep{galdyn08}. While \citet{mukamel08} has noticed that the ergodicity is easily broken for long-range systems, which means that not all the microstates that satisfy the macroscopic constraints  can occur in the long-range systems. From observations and simulations \citep{Battaglia05,navarro10} we know that the velocity dispersion (the ``temperature'') is different at different places of the system, while the velocity distribution at each point is nearly Gaussian \citep{john11}; besides, the force on the particle is mainly determined by the large scale structure of the self-gravitating system \citep{galdyn08}, which may indicate that the particles in a micro volume $\dd^3\textbf{x}$ can be treated as being free. Based on above conditions, we assume that the phase space distribution can be written as
\begin{equation}
\label{local-max}
\overline{f}\propto e^{-\frac{v^2}{6\sigma(r)^2}},
\end{equation}
where $\sigma^2=P/\rho$ and $P$ is the effective pressure defined in \cite{kh11}:
\begin{equation}
\label{eq:peff}
\frac{\dd P} {\dd r} = \frac{\dd p_r}{\dd r} + 2\beta_1 \frac{p_r}{r}.
\end{equation}
$\beta_1$ is the parameter of velocity anisotropy, and $p_r=\rho\sigma_r^2$. We think that eq.(\ref{local-max}) may mean that the ergodicity is broken for the whole system but may still set up locally for arbitrary micro volume $\dd^3\textbf{x}$, because if the system's ergodicity can set up, we can derive the Maxwellian velocity distribution from statistical mechanics, while eq.(\ref{local-max}) means that the Maxwellian velocity distribution only can set up in the micro volume $\dd^3\textbf{x}$. Then from the form of Boltzmann entropy we can obtain (also see section 4.3 of \citep{he11} for the calculations):
\begin{equation}
\label{entropy}
S = -\frac{k}{m}\int\overline{ f}\ln \overline{f}\dd\tau=4\pi \frac{k}{m}\int_0^{\infty} \rho\ln(\frac {P^{3/2}} {\rho^{5/2}}) r^2 \dd r,
\end{equation}
where $\rho$ is the space density, $k$ is the Boltzmann constant, and $m$ is the mass of single particle. Notice that \cite{white87} also studies the elliptical galaxy (which is commonly treated as being collisionless) by using the ideal gas's entropy which is almost the same as eq.(\ref{entropy}). We think that eq.(\ref{local-max}) is applicable both for the systems because eq.(\ref{local-max}) depends only on $\sigma^2$  which can exist even that the two-body relaxation can be neglected or not; and whether the two-body relaxation can be neglected or not, when the system evolves slowly and is close to static, we can both obtain \citep{bett83,galdyn08}
\begin{equation}
\label{hdr}
\frac{\dd P}{\dd r} = -\rho\frac{\dd\Phi}{\dd r}.
\end{equation}
But the difference is that if the two-body relaxation can not be neglected, the $\overline{f}$ in the eq.(\ref{local-max}) can be fine-grained; while for the collisionless system, the Jeans theorem \citep{galdyn08} does not allow the form of eq.(\ref{local-max}), so we need to let $\overline{f}$ to be coarse-grained, and another reason to do this is that $S$ can not evolve if $S$ is a functional of the fine-grained phase density \citep{mo10}. Then with the constraints of mass and energy
\begin{displaymath}
M=\int_0^{\infty}4\pi\rho r^2 \dd r,
\end{displaymath}
\begin{equation}
 E=-2\pi G\int_0^{\infty}\rho(r)m(r)r\dd r=-6\pi\int_0^{\infty} Pr^2\dd r,
\end{equation}
and the differential constraint provided by eq.(\ref{hdr}), we can calculate the entropy's extremum which requires
\begin{equation}
\label{lagrangian}
\delta S_t=\delta(S/k)-\beta \delta E-\alpha \delta M-\delta\int dr\eta(r)(\frac{\dd P}{\dd r} + \rho\frac{\dd\Phi}{\dd r})=0,
\end{equation}
where $\alpha$, $\beta$ and $\eta(r)$ are Lagrangian multipliers. Notice that because of the definition of $P$ we do not require that the velocity anisotropy to be zero during the process of calculating the extremum of the entropy, while observations and simulations also show the non-zero anisotropy in the outer region. Eq.(\ref{lagrangian})'s solution can be approximately written as an equation of state \citep{kh11}
\begin{equation}
\label{eos}
\rho = m\beta P + \alpha P^{n},
\end{equation}
Where n=3/5 or 4/5. The second order variation of $S_t$ is
\begin{equation}
\label{2dervar}
\begin{array}{ll}
\delta^2S_{t} \simeq \displaystyle\frac{1}{2} \int\big[(4\pi+1)\beta\frac{G}{2r^2})(\delta M)^2 \\
\displaystyle-\frac{10\pi r^2}{m\rho}(\delta\rho)^2 -\frac{6 \pi r^2\rho} {mP^2} (\delta P)^2\big] \dd r,
\end{array}
\end{equation}
which indicates that $S_t$ is a saddle point ($\beta>0$). This agrees with the conclusion of \cite{hk11a}.  We can understand it by that based on the broken ergodicity the equilibrium state of self-gravitating system corresponds to the time when the number of microstates is an extremum and does not need to be a maximum , so $S$ is allowed to be not a maximum. Notice that the existence of $\alpha$ ensures that the system can have finite mass and finite energy. High $\alpha$ means small extent (radius) of the system and High $\beta$ requires a small central core (see Fig.~\ref{density}).
\section{Thermodynamics of self-gravitating systems}
\label{3}
In the classical thermodynamics of self-gravitating systems, because of the virial theorem, $2K+W=0$ where $K$ and $W$ are the kinetic and potential energy respectively, $E=K+W=-K$. While people define the system's temperature by $K=3NkT/2$ where $N=M/m$, so
\begin{equation}
\label{ca}
E=-\frac{3}{2}NkT,{\hskip 5mm} C_V=\frac{\partial E}{\partial T}<0.
\end{equation}
An important conclusion is that the negative capacity can cause the gravothermal catastrophe because of the principle of maximum entropy \citep{galdyn08}. However, here we argue about the definition of the temperature $K=3NkT/2$ which has not been proven to be applicable to self-gravitating systems: from eq.(\ref{ca}) we find that $E\propto N$ for fixed $T$, which is not consistent with the non-extensity of long-range systems \citep{salaw00}; although we show that the heat capacity is negative, but it is still positive when the density contrast between the center and the outer part is small enough \citep{galdyn08}. Based on above, we may need to reconsider the thermodynamics of self-gravitating system.

According to the method of statistical mechanics, in the eq.(\ref{lagrangian}) we can identify
\begin{equation}
\beta=1/kT.
\end{equation}
This means that we have found a self-gravitating system's thermodynamic equilibrium state characterized by the temperature $T$, which is a very different conclusion from all previous works. Then a natural question is how to reach this equilibrium state, which will require us to consider the capacity. In the following, we first take $n=3/5$. From figure.~\ref{density}, we know that commonly eq.(\ref{eos}) can be approximated as $kT\rho/m=P$, which will be substituted into the second term of eq.(\ref{eos})
\begin{equation}
\label{eos2}
P=\frac{kT}{m}(\rho-\alpha(\frac{kT}{m})^{\frac{3}{5}}\rho^{\frac{3}{5}}),
\end{equation}
Which is reminiscent of the van der Waals equation that is to describe the non-ideal gas with the interactions among molecules:
\begin{equation}
\label{vander}
(p+\frac{n^2a}{V^2})(V-nb)=NkT,
\end{equation}
where $a$ and $b$ are related to the attractive and repulsive forces respectively. If we take $\rho=Nm/V$ and set $b=0$, we find that eq.(\ref{vander}) will be very similar to eq.(\ref{eos2}), which may further support the rightness of eq.(\ref{eos2}). Multiplying eq.(\ref{eos2}) with $-3/2$ and then integrating the both sides of the equation over the whole volume, we can get
\begin{equation}
E=-\frac{3}{2}kT[N - \alpha (kT)^{\frac{3}{5}}A], A=4\pi\int_0^{\infty} r^2(\frac{\rho}{m})^{3/5}\dd r.
\end{equation}
Notice that here $E$ is not proportional to $N$, which just is the requirement of long-range systems. Bounded systems satisfy $E<0$, which requires that
\begin{equation}
\alpha(kT)^{\frac{3}{5}}<\frac{N}{A}.
\end{equation}
Then we find that the heat capacity is
\begin{equation}
C_V=\frac{\partial E}{\partial T}=\frac{3}{2}k[\frac{8}{5}\alpha (kT)^{\frac{3}{5}}A-N],
\end{equation}
which can be positive or negative. However, according to our results the self-gravitating system can always approach the thermodynamical equilibrium whether the capacity is positive or not, which will be analyzed as the following:

(1). when $C>0$, we can easily obtain
\begin{equation}
\frac{5N}{8A}<\alpha(kT)^{\frac{3}{5}}<\frac{N}{A}.
\end{equation}
For a given value of $\alpha$, $C>0$ is caused by high temperature, then from eq.(\ref{2dervar}) we know that the $(\delta M)^2$ term can be neglected, so the entropy is maximized and the thermodynamical equilibrium can be approached. This also can be understood by that when $T$ is high the self-gravity of our system is very week so that we can treat the system as an ideal gas. When $T$ tends to be infinite, the density profile is a homogeneous distribution truncated at a finite radius, just as shown in figure.\ref{density}. This truncation is caused by the existence of the $\alpha P^{3/5}$ term, which shows us that the existence of $\alpha$ just is equivalent to add an insulting shield to protect the particles from escaping and to ensure of the conservation of the total energy of the system.

(2). $C<0$ requires that
\begin{equation}
0<\alpha (kT)^{\frac{3}{5}}<\frac{5N}{8A}.
\end{equation}
In this case, the classical thermodynamics tells us that the system will become instable and result in gravothermal catastrophe. However, in our new thermodynamics, the entropy $S$ is a saddle point and is not maximized, which means that the second law of thermodynamics is not valid for self-gravitating systems. We first show an extreme case: when $T\rightarrow0$, from eq.(\ref{2dervar}) we know that the  $(\delta M)^2$ term becomes so large that $S$ is minimized, so considering the Clausius statement of the second law of thermodynamics, the heat now is allowed to spontaneously be conducted from low temperature system to high temperature system (In logics we know that if the original proposition is correct, then its converse-negative proposition will also be correct. While current original proposition can be that ``if the second law of thermodynamics is correct, then the entropy of the equilibrium state will be maximized'', its converse-negative proposition is ``if the entropy of the equilibrium state is not maximized, then the second law can be not correct, i.e. the heat now is allowed to spontaneously be conducted from low temperature
to high temperature''), and the thermodynamical equilibrium can still be approached although $C<0$. We also notice that in this case the density profile approaches the singular isothermal sphere and self-gravity becomes very strong, so the self-gravity is the origin of the negative heat capacity, which is consistent with previous study \citep{lyn68}. In the common cases the entropy is a saddle point, and the thermodynamical equilibrium can be approached based on above analysis.

Before comparing our results with observations and simulations, we first talk about the value of $n$. In \cite{kh11} we have said that $n=3/5$ in fact is to treat $\Phi=0$ for large enough $r$, so $\Phi=-GM/r+GM/R$ for $r<R$ where $R$ can be the radius of the system, and $\Phi$ will be zero for $r>R$; while $n=4/5$ is to treat $\Phi=-GM/r$ even for very large $r$. Evidently for observations and simulations the latter one is more accurate than the former one, which also can be seen in Fig.\ref{albadafig}.

\section{Discussions for results of observations and simulations}
\label{4}
In \cite{kh11} we have showed that eq.(\ref{eos}) is very consistent with observations. It can be approximated as truncated cored isothermal sphere which just is the King model \citep{king66} that describes the density distributions of globular clusters. Besides, although the surface brightness profiles of elliptical galaxies are described by the $R^{1/4}$ law, some standard elliptical galaxies, such as NGC 3379, still have a central core, which makes eq.(\ref{eos}) more consistent with their surface brightness than $R^{1/4}$ law \citep{kh11}.

However, some observations \citep{king86} and simulations \citep{miller04} have shown that many globular clusters are ''core collapsed'' which commonly is treated as an evidence of gravothermal catastrophe but can be stopped by the binary stars whether they are primordial or produced by gravothermal catastrophe \citep{galdyn08}. But our understanding is that because of not validity of principle of maximum entropy the gravothermal catastrophe may not exist: the density profile of core collapsed clusters \citep{galdyn08} tends to be the singular isothermal sphere, so according to the analysis of above section the initial conditions (mass,energy,etc) of these clusters make them have strong self-gravity so that their final temperature $T\rightarrow0$ (notice that the these clusters are closer to the galaxy center, which means that they can survive under the strong tidal force of the galaxy, so it is reasonable to believe that they have strong self-gravity), and they are just approaching or have reached the thermodynamical equilibrium. Evidently our explanation does not contradict with the observations and simulations that may manifest the gravothermal catastrophe, but there is another result that may more support our theory: the gravothermal catastrophe will occur once the density contrast between the outer and the inner parts is larger than 708.61 \citep{lyn68}, while \cite{king86} finds that there are a number of globular clusters that have very dense core and very short relaxation times, but they seem not suffering core collapse and are well fitted by the King models.

\begin{figure}
\centerline{\includegraphics[width=\columnwidth]{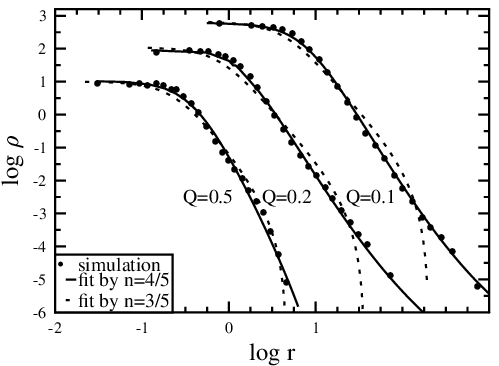}}
\caption{\citet{albada}'s results fitted by eq.(\ref{eos}). $Q=2K/W$.}
\label{albadafig}
\end{figure}

Then we should compare our results with the simulations with isolated boundary conditions. \cite{albada} simulates the dissipationless collapse of galaxies and their figure.~4 shows that for the initial uniform spheres the final density profiles do not have a universal shape, which is thought to be caused by the galaxies' initial collapse factor $2K/W$. But in our view, the results of \cite{albada} just perfectly confirm the rightness of eq.(\ref{eos}): its all kinds of density profiles still can be exactly fitted by eq.(\ref{eos}) with different choices of $\beta$, $\alpha$ and $\rho(0)$ especially when $n=4/5$, which is shown in Fig.\ref{albadafig}. So we conclude that the non-university of the final space density may be not caused by different initial collapse factor but is caused by different total masses and total energies of the simulated systems.
\begin{figure}
\centerline{\includegraphics[width=\columnwidth]{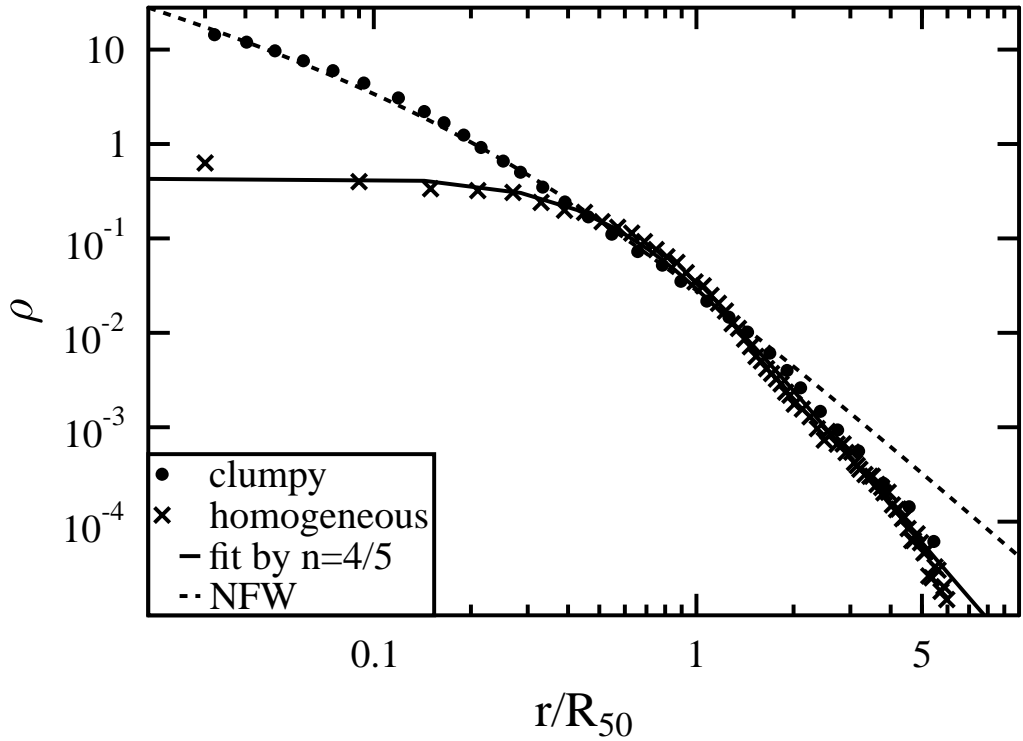}}
\caption{Density profiles of dissipationless simulations with different initial conditions, compared with eq.(\ref{eos}) and NFW profile. $R_{50}$ is the radius with half of mass. We think that the deviations from NFW profile at the outer part is caused by that CDM halos are not strictly isolated.}
\label{core-cusp}
\end{figure}

Until now, our results are consistent with observations of star clusters and elliptical galaxies and simulations of dissipationless collapse. Evidently our results can not explain the cusp profile which is to describe the density profile of the simulated dark matter halos. But we also notice that: \cite{Roy04} also simulates the above collapse with many kinds of initial conditions (but all with the isolated boundary condition): their figure.~3 -- 8 show us that for most initial conditions, the final density profiles have a central core and also can be exactly described by eq.(\ref{eos}); but for the clumpy initial distribution, the cusp can be easily produced and the final central density is better fitted by eq.(\ref{nfw}). All above have been shown in Fig.\ref{core-cusp}. Besides, if the initial density slope is very steep, a cusp will also be produced, but not as quickly as the clumpy case. While \citet{white98} that performs the cosmological simulation has ever suggested that the cusp profile is a result of hierarchy of clustering by mergers of the smaller clumps to the bigger clumps. However, \citet{moore99} has almost rejected this possibility by the simulations of warm dark matter which can smooth the density perturbation at the galaxy scale. So we can conclude that the cored profile only can appear in the simulations with isolated boundary condition and homogeneous initial condition at the galaxy scale, and any disagreements with one of these two conditions will lead to the cusp profile. Besides, observations of LSB galaxies and others strongly support the cored profile; the similarity solution of the secondary infall model of hierarchy clustering, which predicts that the final density profile depends on the initial density perturbation, has shown us that for the initial density perturbation $\delta_i\sim M^{\epsilon}$, the resulting density slope is
\begin{equation}
\gamma=-\frac{\dd \ln\rho}{\dd \ln r}=\frac{9\epsilon}{1+3\epsilon},\hskip6mm  \epsilon>0
\end{equation}
which is a cusp \citep{Nusser01}; our results about the thermodynamical equilibrium state of an isolated self-gravitating system suggest the cored one. If we neglect the interactions between baryons and dark matter halos, do these evidences indicate that the galaxies formed with more isolated boundary condition and more homogeneous initial condition?

\section{Conclusion}
In conclusion, if we assume that the final state of self-gravitating system is the state when the number of microstates (so the entropy) is an extremum, which does not contradict with the ergodicity breaking, we can find a different thermodynamics of self-gravitating system. The equation of state of the equilibrium state is similar to the van der waals equation, and the final density profile is determined by $M$ and $E$ which are used to control the value of $\beta$ and $\alpha$.

Our different thermodynamics states that the equilibrium may always be approached. When the temperature is large enough, our thermodynamics can come back to the classical thermodynamics, and the equilibria of self-gravitating system can be described by the isothermal sphere; when the temperature tends to be zero, the heat capacity is negative, but the the entropy also is minimized, so the system may still can be in equilibrium and gravothermal catastrophe may be not necessary to exist. Some observations and simulations that manifest the gravothermal catastrophe can be treated as a special case of $T\rightarrow0$ in our results.

When we pay attention to the simulations of dissipationless collapse, we find our results can exactly fit their results and show that the the non-universal final density distribution may be not caused by the collapse factor. We also discussed about the core-cusp problem, which may cause many effects at current understandings of galaxy formation. Our results and some explanations are really different from previous works, which needs to be further confirmed.

\section*{Acknowledgements}
We are very grateful for Perez's providing me some related data. This work is supported by the National Basic Research Programm of China, NO:2010CB832805.

\end{document}